\input epsf
\magnification=1095
\parskip=3pt plus 1pt minus 1 pt
\font \titlefont=cmr10 at 14.4truept
\font \namefont=cmr10 at 12truept
\font \small=cmr8

\font \twelvebf=cmbx12
\def\singlespace{\baselineskip=\normalbaselineskip}

\newcount\firstpageno \firstpageno=2
\newcount\sectnumber\sectnumber=0
\outer\def\section#1\par{\advance\sectnumber by 1
     \vskip 17pt plus 4pt minus 3pt\goodbreak
     \leftline{{\twelvebf\number\sectnumber\enspace#1}}
     \nobreak\medskip}
\footline={\ifnum\pageno<\firstpageno{\hfil}\else{\hfil
                                                  \rm\folio\hfil}\fi}
\def\today{\number\year\space \ifcase\month\or  January\or February\or
        March\or April\or May\or June\or July\or August\or September\or
        October\or November\or December\fi\space \number\day}
\def\leaderfill{\kern-0.3em\leaders\hbox to 0.3em{\hss.\hss}%
    \hskip0.1em plus1fill \kern-0.3em}
\def\leaderfil{\leaders\hbox to 0.3em{\hss.\hss}\hfil}

\def\frac#1/#2{{\textstyle{#1\over #2}}}
\def\pp{\par\hangindent=.125truein \hangafter=1}
\def\aref#1;#2;#3;#4{\pp #1, {\it #2}, {\bf #3}, #4}
\def\apress#1;#2;#3{\pp #1, {\it #2}, #3}
\def\apre#1;#2{\pp #1, #2}
\def\abook#1;#2;#3{\pp #1, {\it #2}, #3}
\def\arep#1;#2;#3{\pp #1, #2, #3}
\def\et{\hbox{et al.}}
\def\simgt{\mathrel{\raise.3ex\hbox{$>$\kern-.75em\lower1ex\hbox{$\sim$}}}}
\def\simlt{\mathrel{\raise.3ex\hbox{$<$\kern-.75em\lower1ex\hbox{$\sim$}}}}

\let\gsim=\simgt

\def\deg{\ifmmode ^{\circ}
         \else $^{\circ}$\fi}
\def\pdeg{\ifmmode $\setbox0=\hbox{$^{\circ}$}\rlap{\hskip.11\wd0 .}$^{\circ}
          \else \setbox0=\hbox{$^{\circ}$}\rlap{\hskip.11\wd0 .}$^{\circ}$\fi}
\def\arcs{\ifmmode {^{\scriptscriptstyle\prime\prime}}
          \else $^{\scriptscriptstyle\prime\prime}$\fi}
\def\arcm{\ifmmode {^{\scriptscriptstyle\prime}}
          \else $^{\scriptscriptstyle\prime}$\fi}
\newdimen\sa  \newdimen\sb
\def\parcs{\sa=.07em \sb=.03em
     \ifmmode $\rlap{.}$^{\scriptscriptstyle\prime\kern -\sb\prime}$\kern -\sa$
     \else \rlap{.}$^{\scriptscriptstyle\prime\kern -\sb\prime}$\kern -\sa\fi}
\def\parcm{\sa=.08em \sb=.03em
     \ifmmode $\rlap{.}\kern\sa$^{\scriptscriptstyle\prime}$\kern-\sb$
     \else \rlap{.}\kern\sa$^{\scriptscriptstyle\prime}$\kern-\sb\fi}

\let\arcmins=\parcm

\def\,{\thinspace}
\def\uk{\ifmmode \,\mu$K$\else \,$\mu$\hbox{K}\fi}

\def\COBE{{\sl COBE}-DMR}
\def\kms{\ifmmode $\,km\,s$^{-1}\else \,km\,s$^{-1}$\fi}
\def\hMpc{\ifmmode \,h^{-1}$Mpc$\else \,$h^{-1}$Mpc\fi}
\def\fknee{f_{\rm knee}}

\def\deltat{\Delta t}

\singlespace
\rightline{astro-ph/9602009}

\vskip 3pt plus 0.3fill
\centerline{\titlefont DIRECT IMAGING OF THE CMB FROM SPACE}
\bigskip

\vskip 3pt plus 0.2fill
\centerline{{\namefont
Michael A. Janssen$^1$,
Douglas Scott$^2$,
Martin White$^3$,
Michael D. Seiffert$^4$,}}
\centerline{{\namefont
Charles R. Lawrence$^1$,
Krzysztof M. G{\'o}rski$^5$,
Mark Dragovan$^3$,
Todd Gaier$^4$,}}
\centerline{{\namefont
Ken Ganga$^6$,
Samuel Gulkis$^7$
Andrew E. Lange$^6$,
Steven M. Levin$^1$,
Philip M.  Lubin$^4$,}}
\centerline{{\namefont
Peter Meinhold$^4$,
Anthony C. S. Readhead$^8$,
Paul L. Richards$^{9}$, \&
John E. Ruhl$^4$}}

\vskip 3pt plus 0.3fill

{\narrower
\centerline{ABSTRACT}
\baselineskip=15pt

Fundamental information about the Universe is encoded in anisotropies of 
the Cosmic Microwave Background (CMB) radiation.  To make full use of 
this information, an experiment must image the entire sky with
the angular resolution, sensitivity, and spectral coverage necessary to
reach the limits set by cosmic variance on angular scales
$\simgt10^\prime$.  Recent progress in detector technology allows this to
be achieved by a properly designed space mission that fits well within the
scope of NASA's Medium-class Explorer program.  An essential component of the
mission design is an observing strategy that minimizes systematic effects
due to instrumental offset drifts. 
The detector advances make possible a ``spin chopping'' approach has significant
technical and scientific advantages over the strategy used by COBE,
which reconstructed an image of the sky via inversion of a large matrix of
differential measurements. The advantages include
increased angular resolution, increased sensitivity, and simplicity of
instrumentation and spacecraft operations.  For the parameters typical of
experiments like the Primordial Structures Investigation (PSI)
and the Far InfraRed
Explorer (FIRE), we show that the spin-chopping strategy produces images
of the sky and power spectra of CMB anisotropies that contain no significant
systematic artifacts.} 

\vskip 4pt

{\narrower
{\it Subject headings:} cosmic microwave background---space
vehicles---methods: observational---methods: data analysis}

\vfill

\centerline{Submitted to {\it -- The Astrophysical Journal}}

\vfill
\hrule
\smallskip

\noindent
{\small 1\
Astrophysics 169-506, Jet Propulsion Laboratory, 4800 Oak Grove Drive, 
Pasadena, CA 91109}

\noindent
{\small 2\
Department of Geophysics \& Astronomy
129-2219 Main Mall, University of British Columbia,
Vancouver, B.C.\ \  V6T 1Z4\ \ Canada}

\noindent
{\small 3\
Enrico Fermi Institute, 5640 S.~Ellis Avenue,
Chicago IL 60637}

\noindent
{\small 4\
Department of Physics, University of California,
Santa Barbara, CA 93106}

\noindent
{\small 5\
Code 685, NASA Goddard Space Flight Center, 
Greenbelt, MD 20771, on leave from Warsaw University Observatory, Poland}

\noindent
{\small 6\
Observational Cosmology 59-33, California Inst. of Tech.,  Pasadena, CA 91125}

\noindent
{\small 7\
MS 180-703, Jet Propulsion Laboratory,  4800 Oak Grove Drive, 
Pasadena, CA 91109}

\noindent
{\small 8\
Astronomy 105-24, California Inst. of Tech., Pasadena, CA 91125}

\noindent
{\small 9\
Department of Physics, University of California,
Berkeley, CA 94720}
\eject

\section INTRODUCTION\par

The study of the Cosmic Microwave Background (CMB) entered an exciting new
phase with the detection of large scale anisotropies by the \COBE\ (Smoot et
al.~1992). Degree-scale fluctuations have now been detected by several ground-
and balloon-based experiments (Cheng \et\ 1995; de Bernardis \et\ 1994;
Ganga \et\ 1994; Gundersen \et\ 1995;  Hancock \et\ 1994; Netterfield \et\ 1995;
Piccirillo \et\ 1996;  Ruhl \et\ 1995; Tanaka \et\ 1996).
Full realization of the enormous scientific potential of the CMB, however,
will require a high-sensitivity, high-angular-resolution study of a large
fraction of the sky from a single, stable platform. Due to dramatic advances
in detector sensitivity and stability  over the
last decade, such a study is now feasible, and several groups have devoted
substantial effort to developing plans for a second-generation space experiment
capable of extracting the maximum available cosmological information from the
CMB.

An important element in the design of a CMB experiment is the scan strategy,
which in addition to specifying how different parts of the sky are observed
must provide extremely good control of potential sources of systematic errors. 
The aforementioned advances in detector technology coupled with recent
advances in mission design allow a direct-imaging strategy that has been adopted
by two proposed space missions: the Primordial Structures Investigation (PSI)
and the Far-InfraRed Explorer (FIRE). This strategy has many advantages over
the differential chopping scheme used by \hbox{\COBE}.
The purpose of this paper is to demonstrate,
through analytical calculations and
simulations, that this  direct-imaging strategy can produce images of the sky
containing no significant systematic artifacts, especially those due to
instrumental fluctuations with a $1/f$ spectrum.  We begin by introducing the
scan strategy and its advantages.

PSI and FIRE each propose to use a spinning spacecraft located near the
Earth-Sun $L_2$ point, with the spin axis pointed at the Sun and the
boresight of a single telescope at right angles to the spin axis.
The individual beams of a focal plane array of detectors sweep out great
circles through the ecliptic
poles. The spin axis is precessed either in small steps (6\arcmins67 nine times
daily for PSI) or continuously (FIRE) to maintain Sun-pointing. The output of
the detectors is sampled rapidly ($\gsim3.5$ times per beam) around each
circle.  An image of the whole sky is built up steadily over 6~months as the
Earth and the spacecraft orbit the Sun.  Since each great circle is scanned many
times in succession, instrumental drifts or offsets on timescales long
compared to the spin period are easily removed.  The north and south ecliptic
pole regions tie any given great circle to all others.

In contrast, the differential chopping scheme used by \COBE\  (Boggess et
al.~1992) measures differences in power between two horn antennas or telescopes
pointed at widely-separated positions on the sky. The horns spin around a bisecting
axis, which itself is steered around the sky at a slower rate.  After
essentially the entire sky has been scanned the image is constructed by
inverting a large matrix of temperature differences (Janssen \& Gulkis~1992). 
The differential chopping strategy reduces the effects of gain drifts in
detectors as well as time-varying instrumental or atmospheric offsets, but at
the price of decreased sensitivity and increased complexity of instrumentation
and observing strategy.  Historically this approach
has been required when stability
could be achieved in no other way.  A variant of this strategy has been chosen
for the Microwave Anisotropy Probe (MAP: Wright, Hinshaw, \& Bennett 1995),
another proposed second-generation CMB mission.  

The spin-chopping strategy used by PSI and FIRE has many advantages,
either by itself or as an  facilitating factor in the mission design.  A
complete discussion of these advantages would require full specifications
of the spacecraft and mission designs, which are beyond the scope of the
paper, but we introduce sufficient detail here to motivate the
discussion that follows.  The main advantages are:

\item{(1)}Simplicity of the hardware.
Only a single telescope is required, and the complications of internal switches or
correlation receivers are avoided.

\item{(2)}Superior angular resolution.
A single-telescope design allows the largest possible aperture within the size
limitations of the launch vehicle.  Detailed theoretical investigation of CMB
anisotropies shows unequivocally that even modest increases in angular resolution in
the critical sub-degree region can have dramatic scientific benefits.

\item{(3)}Increased sensitivity.
The low instrument temperatures that can be achieved passively with the
Sun-pointed spin axis (next item) lead directly or indirectly to colder and
more sensitive detectors. Moreover, an unswitched system has a factor of
$\sqrt{2}$ greater sensitivity than a switched system that differences pairs of
pixels on the sky (Janssen \& Gulkis, 1992).

\item{(4)}Extremely benign and stable thermal conditions.  
With the spin axis pointed at the Sun (and $<5^\circ$ from the Earth) the
drivers for spin-synchronous temperature changes are essentially eliminated. A
solar panel at one end of the spacecraft normal to the spin axis shades the
rest of the spacecraft from the Sun and Earth at all times.  Passive cooling of
the instrument end of the spacecraft to below $60\,$K is readily achieved.

\item{(5)}A direct image of the sky is formed incrementally as it is observed.
One does not have to wait until a large fraction of the sky (for \COBE\ almost
the entire sky) is observed to form an image.  This significantly reduces the
science risk of an early mission failure, and allows immediate investigation of
mission performance and potential systematic errors at the microkelvin level.

\item{(6)}It gives a direct measure of $1/f$ noise in the instrument, since
each great circle is ``closed.''  The differential strategy is vulnerable to
drifts in the switched system offset.  While this was relatively easy to handle
in \hbox{\COBE}, the problem becomes more severe at higher angular resolutions
and sensitivities where the beams are not swapped precisely on the sky in the
measurement of each pixel pair.  Also, thermal gradients are more difficult to
control with two large telescopes than with two small horns.

\item{(7)}Data from different parts of the sky are independent.  There is no
aliasing of signal from high foreground regions such as the Galactic plane into
foreground-free regions as there may be in differential chopping schemes
(see also Lineweaver \et\ 1994).  Similarly, the effects of transients, either in the
instrument or on the sky (e.g., variable or moving sources such as planets) are
localized, and the large number of scans over a given great circle provides many
cross-checks on a pixel-by-pixel basis.  

\item{(8)} Simplicity of spacecraft operations.  The almost stationary spin
axis simplifies attitude control and ground communications.

\smallskip

These advantages would be lost if medium-timescale instrumental
fluctuations that could not be ``chopped out'' by the spacecraft spin
introduced serious systematic errors.  Such fluctuations might be caused by,
for example, detector gain variations or temperature changes in the
optical system with a $1/f$ spectrum.  Naively one might expect that the
detectors would have to have essentially no such fluctuations on timescales
shorter than the spin period. In this paper we derive analytical expressions
for the effect of $1/f$ noise on sky images obtained with the scan strategy
described.  We also verify the
analytical results with simulations of sky images at very high resolution,
using detector characteristics measured in the laboratory for both transistor
amplifiers (for PSI) and bolometers (for FIRE). We show that the naive
expectation that detector fluctuations must be on timescales longer than the
spin period is far too restrictive, and that a spin-chopping mission in the
benign $L_2$ environment with recently developed and characterized transistor
amplifiers or bolometers can realize all of the above advantages.

\section THEORETICAL ANALYSIS

The spacecraft spin acts as an ideal switch or chop to suppress slowly varying
drifts and offsets; however, the effects of instrumental drifts on timescales
less than the spin period cannot be suppressed completely by the spin. In an
instrument based on either indium phosphide high-electron-mobility transistor
amplifiers (PSI) or AC-biased bolometers operated from a temperature regulated
heat sink (FIRE), these drifts are measured to have a $1/f$ spectrum dominated
by transistor gain fluctuations.  

In this section we derive analytic expressions for the effect of such drifts
on an image of the sky on a single scan circle (combining different scan circles
is discussed in the next section).  The main effect turns out to be an increase in
the rms noise in the image that depends on the spacecraft spin rate and on the
frequency at which the $1/f$ and the ``white'' noise have the same amplitude, known
as the $1/f$ knee frequency. 

Assume a time-varying signal $T(t)$ due to stationary noise that is
characterized by the noise power spectrum $S(f)$, where $f$ is frequency.
The signal is measured as a uniform series of contiguous averages over
integration time $\deltat$,
$$T_i = \int^{t_0+i\deltat}_{t_0+(i-1)\deltat} T(t)\ dt,\eqno(1)$$
where $t_0$ is some arbitrary reference time.
We wish to determine the variance
$$\sigma^2(j-k)={1\over2}\left\langle(T_j-T_k)^2\right\rangle\eqno(2)$$
in terms of the time-lag index $j-k$, the integration time $\deltat$, and the
noise power spectrum $S(f)$.  Angle brackets denote ensemble averages.
Defining $n\equiv j-k$ and using the stationarity property of the noise,
this equation may be rewritten as
$$\sigma^2(n)=\left\langle T_j^2\right\rangle - \left\langle T_j
T_k \right\rangle.\eqno(3)$$
Using Eqn.~(1) and
$$\left\langle T(t)T(t-\tau)\right\rangle
 = \int_0^{\infty} S(f) \cos\left( 2\pi f\tau\right)\ df,\eqno(4)$$
we find
$$\sigma^2(n) = {2\over \deltat^2} \int_0^{\infty}
 S(f)\ {\sin^2(\pi f\deltat)\over(\pi f)^2}
 \ \sin^2(n\pi f\deltat)\ df.
\eqno(5)$$

We have measured the detected-power spectrum $S(f)$ of
the InP HEMT amplifiers and bolometers that will be used by PSI (Seiffert \et\
1996; Gaier \et\ 1996b) and FIRE, respectively.  For both, $S$ is well-represented by
$$S(f) = a + b/f\eqno(6)$$
for which Eqn.~(5) becomes
$$
\sigma^2(n) = {a\over 2\deltat}\left[ 1+{b\over a}\deltat\ \phi(n)\right],
\eqno(7)
$$
where
$$\eqalign{
\phi(n) &= (n-1)^2\ \ln(n-1)\ -\ 2 n^2\ \ln n\ +\ (n+1)^2\ \ln(n+1)\cr
&\to 2\ln n+3\quad\hbox{as}\ n\to\infty.}
\eqno(8)$$
(Convergence is rapid.  For $n\geq10$ the error is $\leq10^{-3}$.)  The
logarithmic behavior of $\phi$, essential in what follows, is illustrated in
Figure~1.  Note that the noise characteristics of many detectors {\it
cannot\/} be represented by Eq.~(6).  For example, the detectors used on IRAS
exhibited strong memory effects, which introduced large offsets whenever a
bright object was crossed.  Such detectors would be quite unsuitable for
spin-chopped observations of the CMB.

\goodbreak
\midinsert
\centerline{
\epsfxsize=13cm \epsfysize=13cm \epsfbox{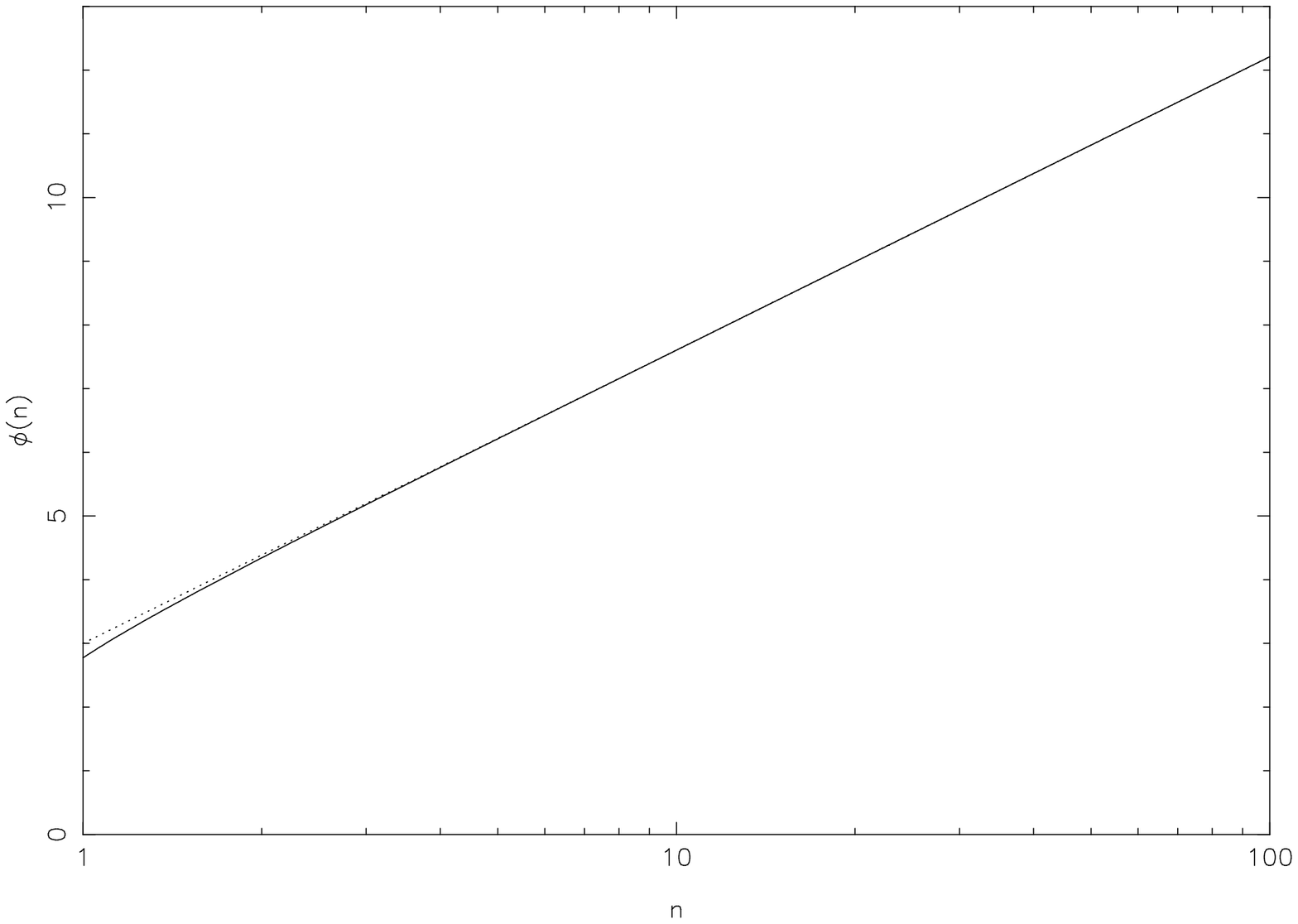}}
\nobreak
\vskip\baselineskip
{\bf Fig.~1}\
$\phi(n)$ from Eqn.~(8).  The dotted line shows the approximation $\phi =
2\ln  n + 3$, which gives an error $\leq10^{-3}$ for $n\geq10$.
\endinsert

The variance $\sigma_0^2$ for a total power radiometer with pure white noise
(i.e., $b=0$ in Eqn.~(6)) is well known to be
$$\sigma_0^2 = {T_s^2 \over B\ \deltat},\eqno(9)$$
where $T_s$ is the system noise temperature, $B$ is the radiometer bandwidth,
and $\deltat$ is the averaging time for a given sample as defined in
Eqn.~(1).  Eqn.~(7) may then be written as
$$\sigma^2(n) = \sigma_0^2\ \bigl[ 1+ \deltat\ \fknee\ \phi(n)\bigr]\eqno(10)$$
where $\fknee\equiv b/a$ is the frequency at which the white noise and $1/f$ noise
contributions are equal. The factor
$$F = \bigl[ 1+ \deltat\ \fknee\ \phi(n)\bigr]^{1/2}\eqno(11)$$
thus gives the increased uncertainty over the case of perfect white noise.

Now consider a scan circle divided into $m$ uniform and contiguous
integrations, and assume that the circle is scanned $N$ times and that the $N$
measurements at each pixel are averaged.  Specifically, for two pixels
$1\leq p<q \leq m$ around the circle, these averages are
$$\eqalign{\overline T_p &= {1\over N}\sum_{i=1}^N T_{p+(i-1)m}\,,\cr
{\rm and}\quad\overline T_q &=
{1\over N}\sum_{i=1}^N T_{q+(i-1)m}\,.\cr} \eqno(12)$$

Let us find the variance
$$\sigma^2(n=q-p,m,N)={1\over2}
\left\langle(\overline T_q-\overline T_p)^2\right\rangle.\eqno(13)$$
Substituting Eqn.~(12) into (13), expanding, and making use of Eqn.~(3),
we obtain
$$\sigma^2(n,m,N) = {1\over N} \sigma^2 (n)\ -\
{1\over N ^2 } \sum_{k=1}^{N-1} (N-k)\ \Bigl\{  2\sigma^2(km)-
  \sigma^2\bigl[km+n\bigr]-\sigma^2\bigl[km-n\bigr]  \Bigr\}.\eqno(14)$$
Using Eqn.~(10), we arrive at
$$\sigma^2(n,m,N) = {\sigma_0^2\over N}\ {\Bigl[1+\deltat \ \fknee\
\Phi\left(n,m,N\right)\Bigr]},\eqno(15)$$
where
$$\Phi(n,m,N) = \phi(n) - {1\over N} \sum_{k=1}^{N-1}
(N-k)\ \Bigl\{ { 2\phi(km)- \phi\bigl[km+n\bigr]
-\phi\bigl[km-n\bigr]  }\Bigr\}.\eqno(16)$$

The factor
$$F(n,m,N) = \Bigl[1 + \deltat\ \fknee\ \Phi(n,m,N)\Bigr] ^{1/2}\eqno(17)$$
gives the increase in image uncertainty in the scan-averaged image over the
white-noise case.  Figure~2 shows $\Phi(n,m,N)$ for typical parameters. The
case $N=1$ consists of only one scan with no scan averaging, so that the factor
$F$ increases simply as the two-point variance.  As $N$ increases $\Phi$
becomes symmetric around the midpoint, reflecting the fact that pixels beyond
the midpoint move closer to the first pixel of the next scan, shortening the
effective time between pixel pairs.  PSI will average 1600 scans before
redirecting the spacecraft spin axis, so that we are practically concerned only
with the asymptotic form for large $N$. FIRE will have $N\gg1$, but much
smaller than for PSI.  Note that for $N\gsim10$, $\Phi(n,m,N)$ depends hardly
at all on $N$, and  $\sigma^2 \propto N^{-1}$.  

\midinsert
\centerline{
\epsfxsize=13cm \epsfysize=13cm \epsfbox{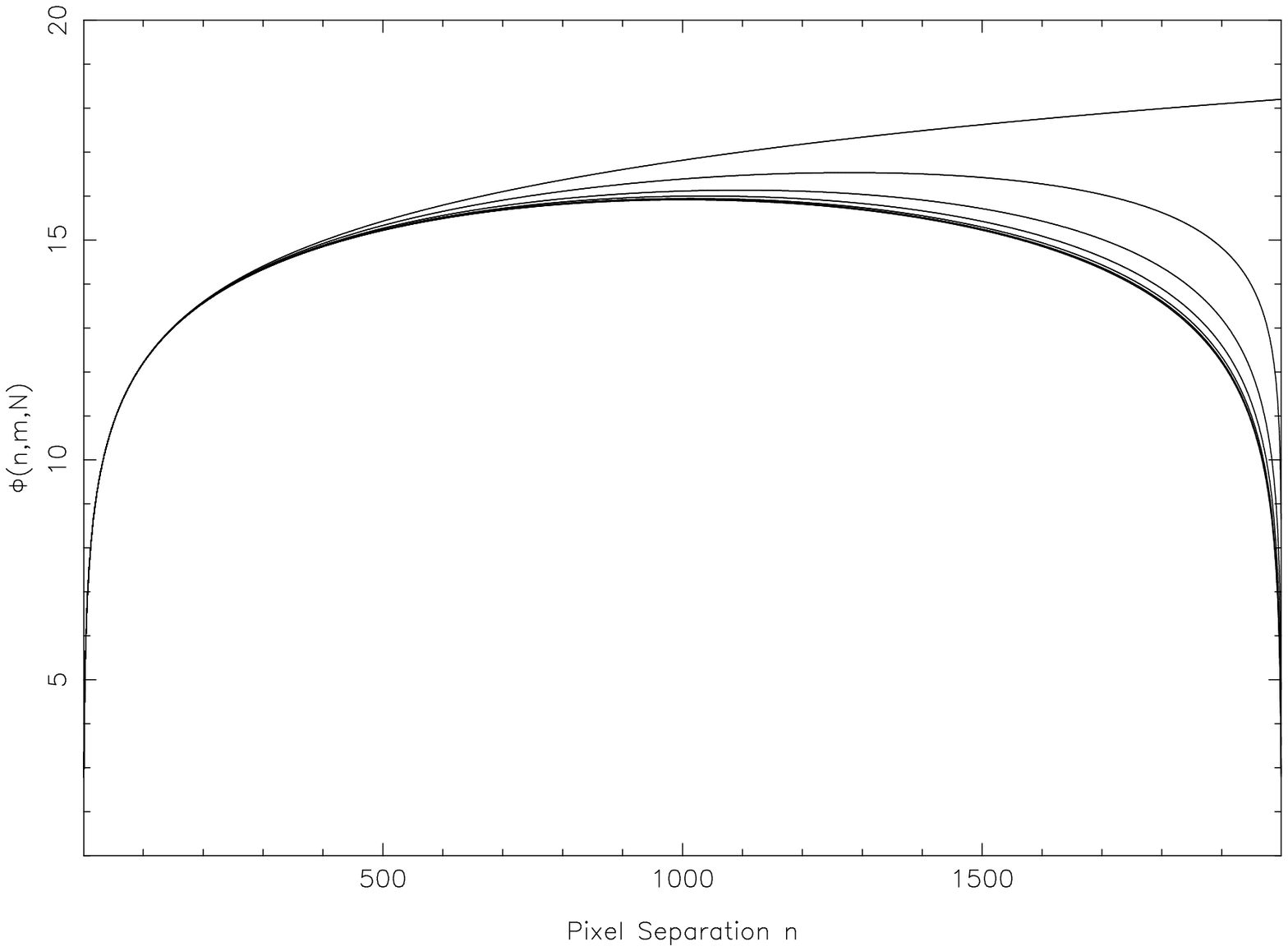}}
\nobreak
\vskip\baselineskip
{\bf Fig.~2}\
$\Phi(n,m,N)$ from Eqn.~(16) as a function of pixel separation $n$ for a scan
circle  consisting of $m = 2000$ pixels.  Seven curves are plotted, for (top to
bottom)  $N=1$, 3, 10, 30, 100, 300, and 1600.  The curves for $N=300$ and
$N=1600$ are almost indistinguishable.  For PSI, $m=1700$ (corresponding to a
12\arcmins6 beam) and $N=1600$.  For FIRE, $m=2700$ (corresponding to an
$8^\prime$ beam) and $N=18$.  For large $m$ the curves are insensitive to $m$.
\endinsert

The uncertainty is a maximum between diametrically opposed pixels if $N$ is
large.  From the above we may write
$$F_{\rm max} (m,N) = F(m/2,m,N) = \Bigl[ 1+\deltat\  \fknee\ \Phi_{\rm max}
(m,N)\Bigr]^{1/2},\eqno(18)$$
$$\Phi_{\rm max} (m,N) = \phi(m/2) - {1\over N} \sum_{k=1}^{N-1}
(N-k)\ \Bigl\{ { 2\phi(km)- \phi\bigl[\left(k+1/2\right)m)\bigr]
-\phi\bigl[\left(k-1/2\right)m\bigr]  }\Bigr\}.\eqno(19)$$
For large $N$ the sum may be approximated as an integral and the term in
$\{\cdots\}$ is approximately $\phi''(x)\sim x^{-2}$.  The sum is thus only
logarithmically dependent on $m$. Numerical evaluation gives $0.871$ for
$N>50$ almost independent of $m$.  Hence we can approximate
$$\Phi_{\rm max} (m) = \lim_{N\to \infty} \Phi_{\rm max} (m,N) \simeq
\phi(m/2) - 0.871.\eqno(20)$$
Further, if $m\simgt 10$, we can simplify this to
$$\Phi_{\rm max} (m) \simeq 2\ln m + 0.743.\eqno(21)$$
The maximum excess noise factor for a typical PSI or FIRE scan circle can
therefore be approximated closely as
$$F_{\rm max} \simeq \bigl[1 + \deltat\  \fknee\ (2\ln m + 0.743
)\bigr]^{1/2}.\eqno(22)$$
$F_{\rm max}$ is plotted in Figure~3 as a function of $\fknee$ for the PSI and
FIRE spin rates. Note that the sensitivity of a differential chopping
instrument corresponds to $F_{\rm max} = \sqrt{2}$.

\midinsert
\centerline{
\epsfxsize=13cm \epsfysize=13cm \epsfbox{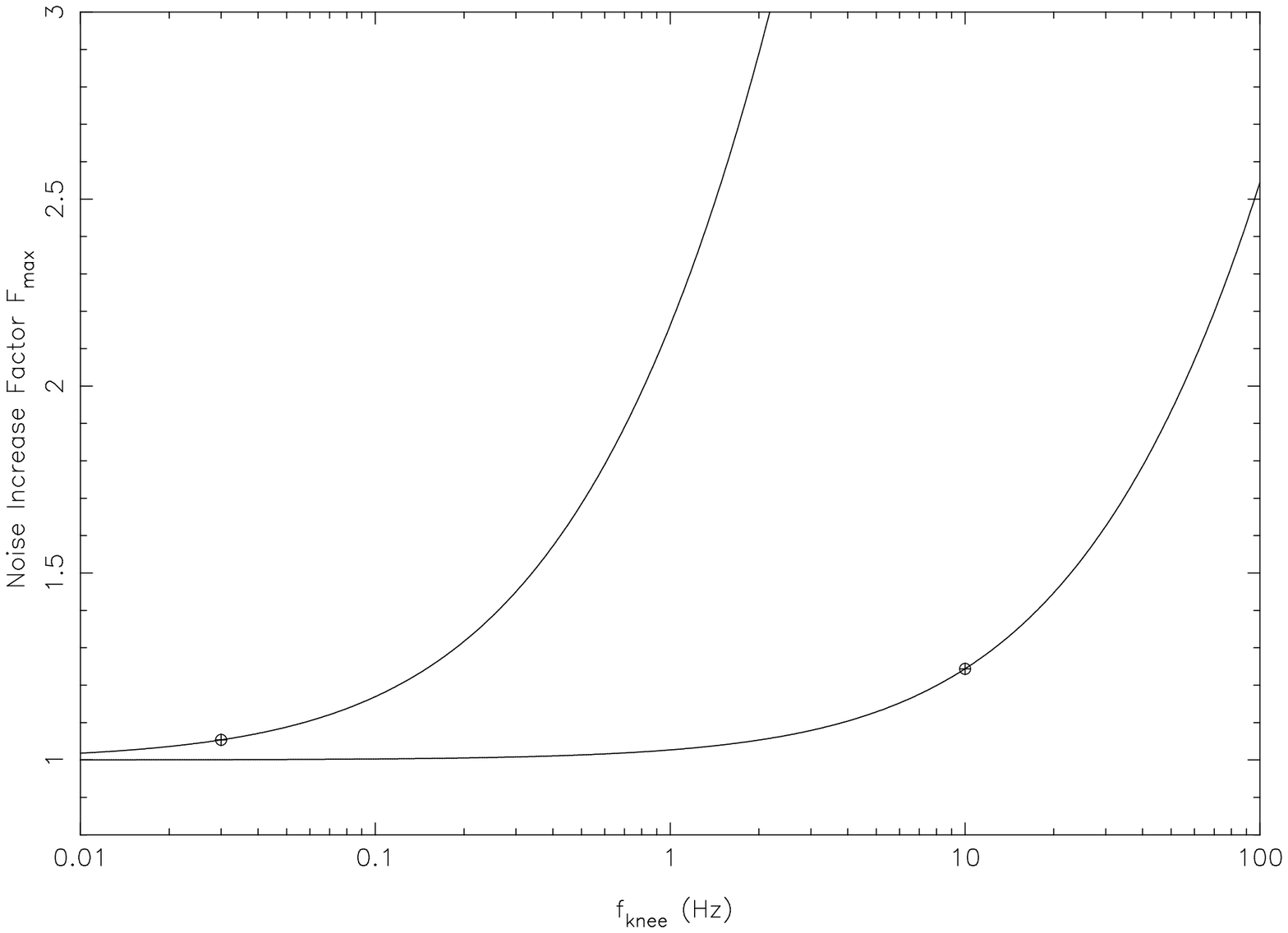}}
\nobreak
\vskip\baselineskip
{\bf Fig.~3}\ 
$F_{\rm max}$ from Eqn.~(22) versus $\fknee$ for the PSI and FIRE spin
rates of 10 and 0.1\,rpm, respectively.  The pixel size for the PSI curve (lower) is
12\parcm6, while that for the FIRE curve is $8^\prime$.  The symbols mark
knee frequencies measured in the laboratory for stabilized InP HEMT amplifiers
for PSI (10\,Hz; Seiffert \et\ 1996; Gaier \et\ 1996b) and bolometers for FIRE
(0.03\,Hz). \endinsert

Based on laboratory measurements, $\fknee\simeq 10$\,Hz for the stabilized
HEMTs used by PSI (Seiffert \et\ 1996; Gaier \et\ 1996b), and $\fknee\simeq 0.03$\,Hz
for the bolometers used by \hbox{FIRE}. These knee frequencies correspond to
increases in the noise over the $S(f)={\rm constant}$ (pure white noise) case of only
about 24\% for PSI and 6\% for \hbox{FIRE}. This is significantly less even than the
minimum increase of noise in a switching system of $\sqrt{2}$.

\section SIMULATIONS\par

The foregoing calculation gives the variance between pixels on a given
scan circle, which does not depend on the mean value. 
The variance between pixels on {\it different\/} scan circles, however, does
depend on the mean values of the two circles.  Since we are
interested only in CMB anisotropies, the mean level itself is unimportant.  On the
other hand, since the mean value of a $1/f$ noise component formally diverges as the
averaging time increases, the mean levels on different scan circles will in general be
different, and if not removed would produce stripes parallel to the scan direction. 

Assume for the moment that the offsets between scan circles have been determined
and removed.  The variance between pixels on the circle is unchanged.  The variance
between random pixels on {\it different\/} circles depends on how accurately
the offset was removed and on the relative phases of the dominant $1/f$ fluctuations
along the two circles.  If we assume that the offset has been subtracted with an
error small compared to $\sigma_{\rm max}\equiv \sigma_0 (F_{\rm max} / N)^{1/2}$,
then $\sigma^2_{\rm max}$ should be an upper bound on the variance calculated
between random pixels on different circles.  

\midinsert
\centerline{
\epsfxsize=13cm \epsfysize=7cm \epsfbox{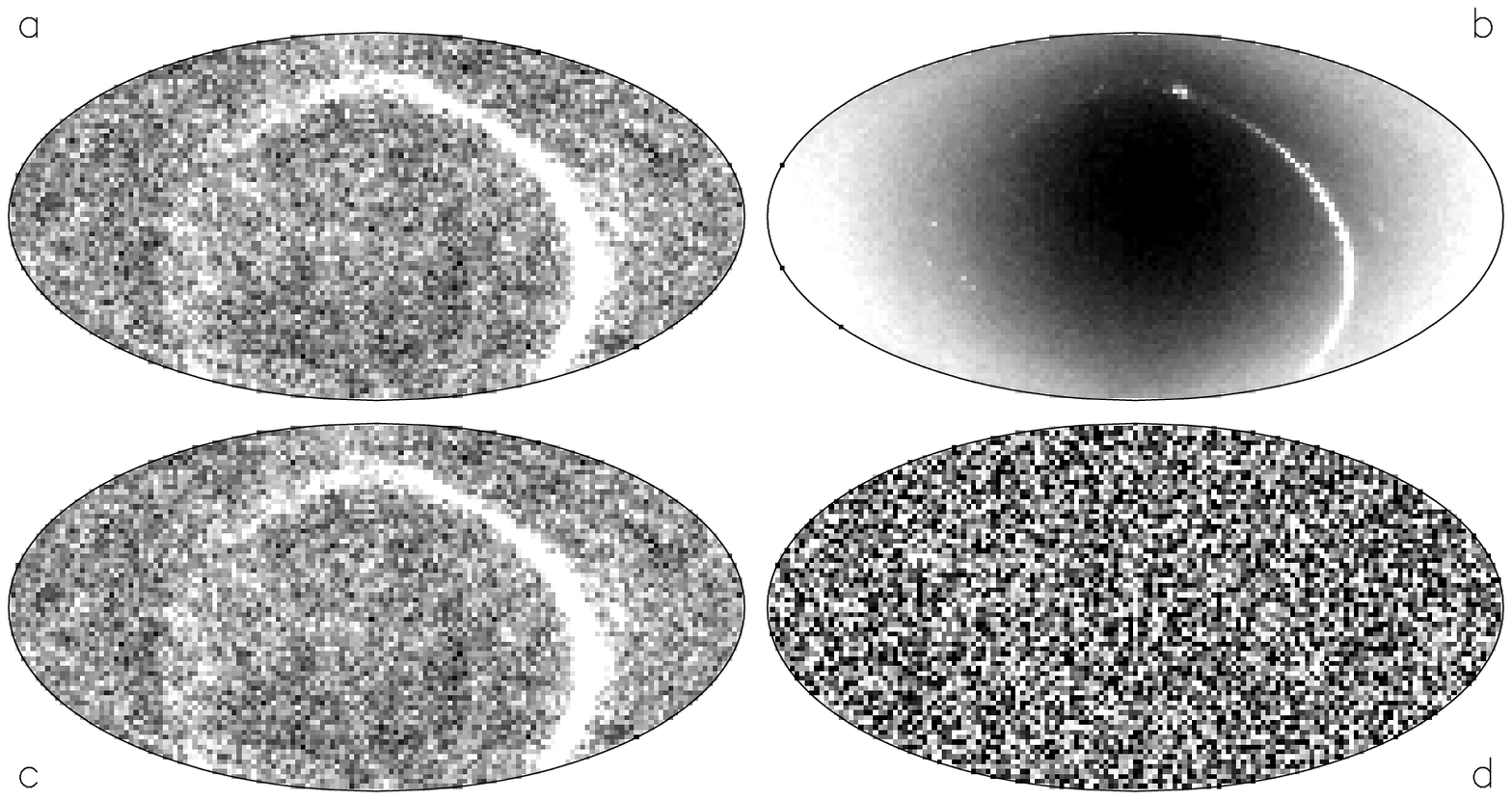}}
\nobreak
\vskip\baselineskip
{\bf Fig.~4}\
All-sky PSI simulation at $90\,$GHz, including both white and $1/f$ noise.
The FIRE simulation at a similar frequency is indistinguishable from this one.
All images are Aitoff projections in ecliptic coordinates,
with $12.6^\prime$ pixels, to match roughly the resolution of the
printer.  (For a standard printer at $600\,$dots per inch, this gives
$1.25\,$dots per pixel, and the image is therefore black and white,
not grayscale.)
\quad {\bf a)}~Assumed sky, using a standard CDM model to generate CMB
fluctuations together with extrapolations of Galactic data taken at other
frequencies.  The stretch is $\pm300\uk$.
\quad{\bf b)}~Same as before, but including instrument noise (white noise +
$1/f$ noise as measured in the laboratory at $90\,$GHz) and the dipole.  
The stretch is $\pm3000\uk$ to accommodate the dipole.
\quad  {\bf c)}~Observed sky, following fitting and removal of the dipole, plus
DC-offset and gradient terms in each great circle.  The stretch is 
$\pm300\uk$.
\quad {\bf d)}~Difference between lower left and upper left images
stretched to $\pm30\uk$, showing the high SNR achieved.  Residual striping
is seen along lines of constant longitude at the level expected from the
calculations in \S~2.  Stripes parallel to the
sides of the paper are caused by uneveness in the printer, not the
experiment! 
\quad {\bf Note:}~the figure above has been drastically reduced in resolution
in order to be a manageable $100\,$k postscript file.  Full resolution and
colour versions of this figure can be found at
{\tt http://astrophysics.jpl.nasa.gov/PSI}.
\vskip 0pt plus 1600fill
\endinsert

To complete our demonstration that the spin-chopping strategy leads to no
significant systematic errors due to medium-timescale instrumental
fluctuations, we now describe simulations of full-sky images that:
1)~demonstrate that offsets between scan circles can be removed; 2)~verify the
conjecture that $\sigma^2_{\rm max}$ is a good estimator of the variance in
the full-sky images; and 3)~show that there are no significant artifacts of
any kind in the images.  

We simulated a CMB signal at $<1'$ resolution assuming a standard Cold Dark
Matter (CDM) model with $h=0.5$, $\Omega_0=1$ and $\Omega_B=0.05$.  To this
cosmic signal we added Galactic emission extrapolated from IRAS and DIRBE
(for the dust) and ground based measurements (for free-free and synchrotron),
as well as the CMB dipole.

A time series of signal data was generated by ``scanning'' the simulated sky
according to the PSI and FIRE prescriptions, to which was added a simulated time
series of noise.  This time stream was binned into the pixel on the sky
corresponding to the spacecraft attitude. The noise time stream was generated
in runs of length corresponding to $\simeq2\times10^6$ pixels.  Since each
great circle contains $\sim6\times10^3$ pixels, this allowed us to wrap 60 (30)
times on each scan circle for PSI (FIRE), and cover $6$ adjacent scan circles
with correlated noise. For FIRE we simulated the actual number of rotations
per great circle, while for PSI we stopped for practicality at $N=60$.  As can
be seen in Fig.~2, we are approaching the asymptotic regime at this $N$, as
varying the number of rotations in the simulations confirmed. By choosing
$N=60$ we were able to obtain coherence in the simulated time stream over
several great circles in the sky to maximize our sensitivity to striping.
Lower resolution simulations with even more correlated great circles showed no
extra cross-scan striping.

\midinsert
\centerline{
\epsfxsize=13cm \epsfysize=8.75cm \epsfbox{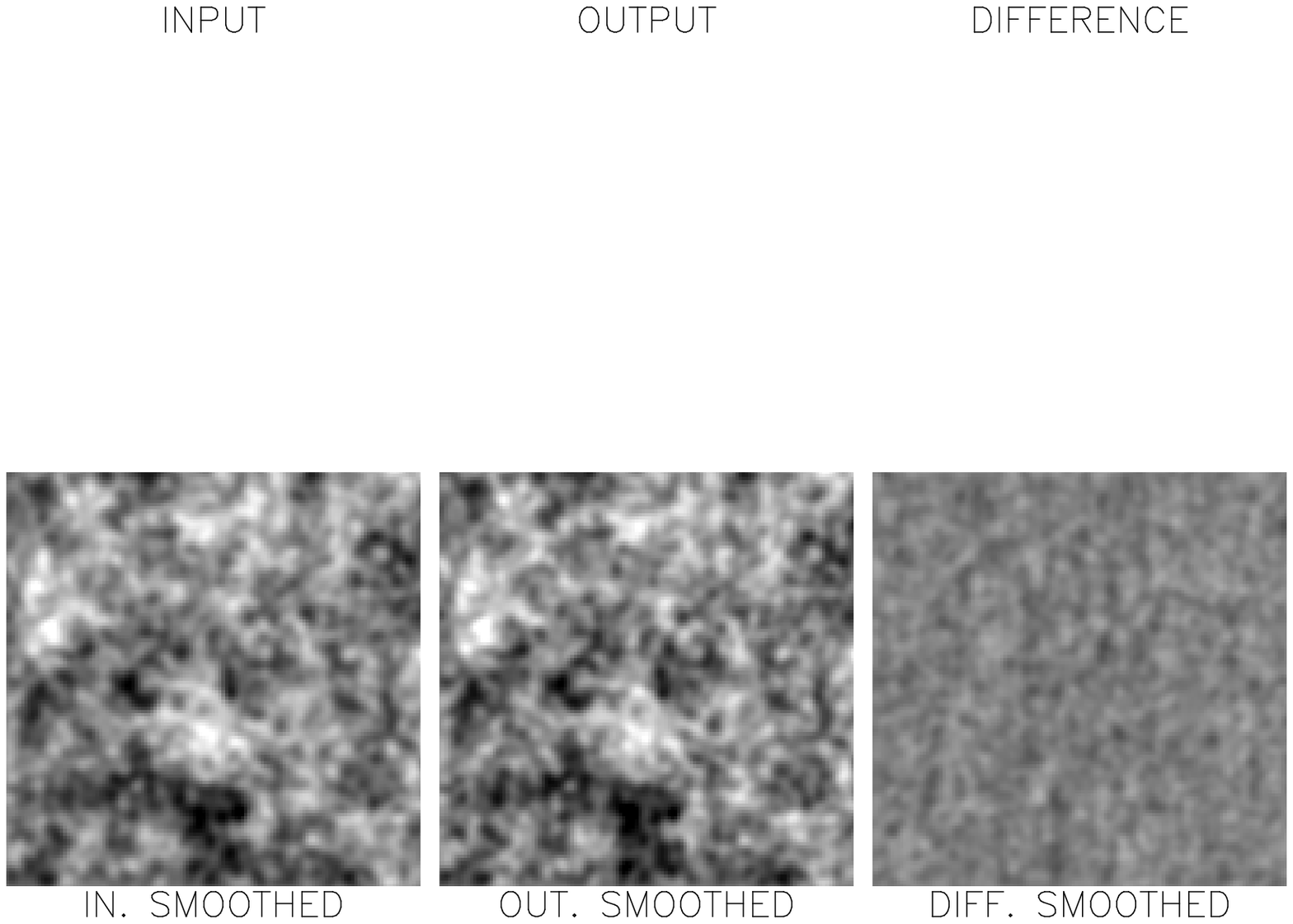}}
\nobreak
\vskip\baselineskip
{\bf Fig. 5a}\ PSI simulation of a small patch, enlarged for clarity. Same as
Figure~4, for a $14^\circ \times 14^\circ$  equatorial region (where
sensitivity and the possible effects of $1/f$ noise are worst).  The pixel size
is  3\arcmins6 (i.e.~$3.5\times$ oversampling).  The saturation level has been
set at $\pm200\,\mu$K for this plot.  The bottom row is smoothed to the
12\arcmins6 beam size.  The scan direction is vertical. \quad {\bf Note:}
Full 6 panel ($8\,$k) version of this figure available at
{\tt http://astrophysics.jpl.nasa.gov/PSI/}.
\endinsert

For both PSI and FIRE, the power spectrum of the simulated noise was derived
from real amplifier data.  These data supported the assumption of a uniform
random phase distribution for the simulated data, allowing us to generate the
time streams in the simulation from the known power spectra.
The noise added into the time stream was scaled relative to the signal map
to the level expected {\it after\/} foreground removal from 2~years of
observation.  This level was determined in separate simulations (not described here)
of foreground removal that used the raw instrumental noise levels (see Gaier \et\
1996a for 90\,GHz performance) and beam sizes expected at each frequency (similar but
preliminary studies are described in Brandt et al.~1995).

For PSI the data stream came from $44\,$GHz tone-stabilized HEMTs, sampled at 500\,Hz
for 500\,s.  We derived a power spectrum that was well-fitted by white noise plus
$1/f$ noise with $f_{\rm knee}=10\,$Hz.  Similar measurements for FIRE bolometers
gave a spectrum well-fitted by white noise plus $1/f$ noise with  $f_{\rm
knee}=0.03\,$Hz.

The $N$ rotations were averaged together to produce one great circle of data.
Offsets between scan circles were removed simply by subtracting the mean on each
circle.  (For FIRE a gradient was also subtracted, to remove the small residual
asymmetry shown in Figure~2 for $N=30$.  In practice this has essentially no
effect on sky power, since the sky must ``close'' around the circle.) 
The result is shown in Figure~4, which looks identical at 600 dots per inch
for either the PSI $90\,$GHz channel or the FIRE $100\,$GHz channel.
Note that time domain drift removal in our configuration is easier than it
was for \COBE\ (e.g.~Bennett et al.~1994), since we reference to the poles
in every scan.

\midinsert
\centerline{
\epsfxsize=13cm \epsfysize=8.75cm \epsfbox{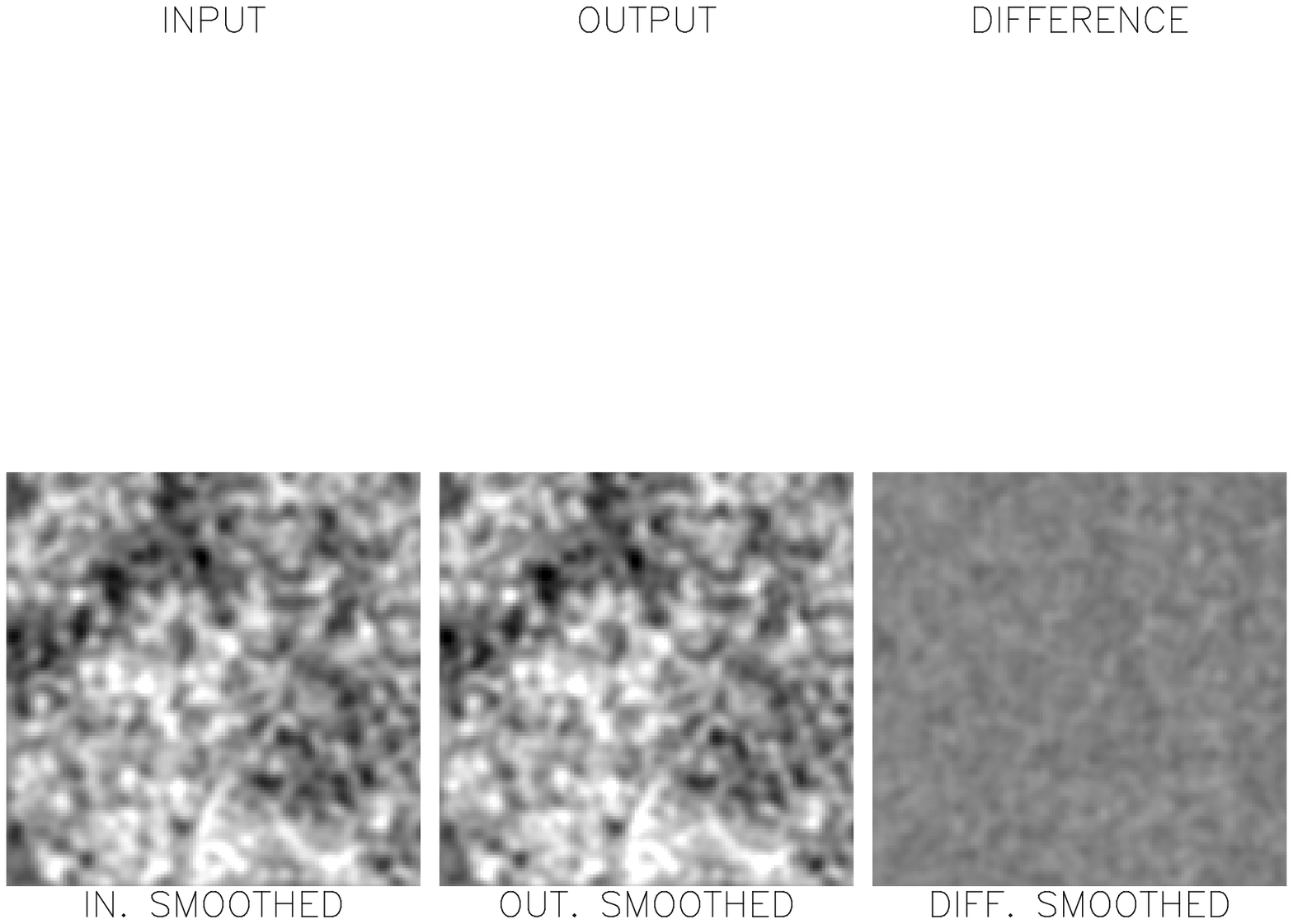}}
\nobreak
\vskip\baselineskip
{\bf Fig. 5b}\ FIRE simulation of a small patch, enlarged for clarity.
Same as before, with a $10^\circ \times 10^\circ$  equatorial region.
The pixel size here is 2\arcmins25$\times$2\arcmins25.  The bottom row is
smoothed to the 8\arcm\ beam size.  The scan direction is vertical.
\quad{\bf Note:}
full 6 panel ($8\,$k) version of this figure available at
{\tt http://astrophysics.jpl.nasa.gov/PSI/}.
\endinsert

Figure~5a shows a $14^\circ\times14^\circ$ patch of sky for the PSI experiment
enlarged for clarity to the point that the printer can show a true gray scale. 
The beam size is 12\parcm6 and the pixel size is 3\parcm6, set by the sampling
rate.  The top row is shown at full pixel resolution; the bottom row is
smoothed to the beam size.  Left to right the panels show the assumed and
observed skies, and the difference.  The signal-to-noise ratio (SNR) is 1.1
even on the {\it pixel\/} scale. There are small residual ``stripes'' along
scan lines (vertical here), exactly as expected from the theoretical
calculations.  An additional destriping step was performed for the PSI small
patch simulations, where the mean of each scan across the patch was subtracted.

Figure~5b shows the same thing for the parameters of the FIRE experiment at
$220\,$GHz, the highest frequency channel at which CMB fluctuations are likely
to be usefully measured.  The beam size is $8^\prime$ and the pixels are
2\parcm25, again set by the sampling rate.  The patch plotted here is
$10^\circ\times10^\circ$, and the panels represent the same things as for the
PSI figure, except that here the lower three panels have now been smoothed by
$8^\prime$. The SNR achieved by PSI and FIRE will allow super-resolution, i.e.,
information can be extracted on angular scales substantially smaller than the beam
size,  and in practice one would make images that over-sample the beam by
significantly larger factors than we have used here.  

To quantify in a different way the effect of the residual striping
along scan lines we have extracted the power spectrum of fluctuations
from our simulated skies.  Simulated full-sky images at the highest achievable
resolution were observed with the PSI or FIRE strategy, including the effects of
noise. From these we calculated the $C_\ell$ spectrum, where $C_\ell$ is the usual
squared amplitude of the average spherical harmonic at multipole $\ell$ (see e.g.,
White, Scott \& Silk~1994). Each input $C_\ell$ is one realization of the
underlying average sky, and hence the power spectra show the effects of cosmic
and sample variance at each multipole.  We assumed for convenience that foregrounds
had been removed over 80\% of the sky with attendant increase in the noise level (see
above).  The other 20\% of the sky (the Galactic plane) was ignored.  In practice
the fraction of the sky contaminated with foregrounds may be somewhat higher.  The
only effect of this would be a corresponding increase in sample variance.

Figure~6 shows the spectrum of multipole moments derived from our input, output, and
difference images.  Figure~6a is for a PSI simulation at $90\,$GHz, with the 12\parcm6
beamsize.  Figure~6b shows the same thing for FIRE, with a beam-size of $8^\prime$.

Each plot has two panels: the upper panel has $\ell(\ell+1)C_\ell$ on the vertical
axis,  so that the plot is power spectrum per logarithmic interval in $\ell$; the
lower panel has $C_\ell$ on the vertical axis, this being the natural way to indicate
noise, since pure white noise is then a horizontal line.  The $C_\ell$'s extracted
from the input image are shown shifted vertically by two orders of magnitude to
separate them from the output image $C_\ell$'s.  Note that both the input and output
power spectra are the underlying theoretical spectra multiplied by the experimental
window function, which is why they drop off faster at high $\ell$ than the more
familiar theoretical curves.  Also note that, as expected, the output power spectrum
is the sum of the input and the noise spectra.  

Four points are worth emphasizing: 1)~no significant artifacts are apparent in the
output sky power spectrum; 2)~the difference image spectrum (i.e., the noise
estimator) is nearly white noise over all angular scales, with some extra power
at large scales where the SNR is largest;  3)~the variance over the sky is
$\sigma^2_{\rm max}$, as expected; and 4)~we expect to be able to determine the
power spectrum efficiently and {\it without smoothing\/} out to $\ell\simeq900$ (for
PSI) and $\ell\simeq1400$ (for FIRE).

These imaging simulations reinforce our analytic calculations.  Not only are
there no significant artifacts in the images, there is none in the power
spectra either.  The increase of the RMS noise over the pure white noise case
is exactly as predicted by Eq.~(22).
Moreover, the simulations demonstrate that it is straightforward to analyze the
time-stream data and produce a two-dimensional image of the sky.  Many refinements
of the method are possible, particularly concerning removal of offsets between
scan circles, but even the simple method we used here works extremely well.

\midinsert
\centerline{
\epsfxsize=13cm \epsfysize=13cm \epsfbox{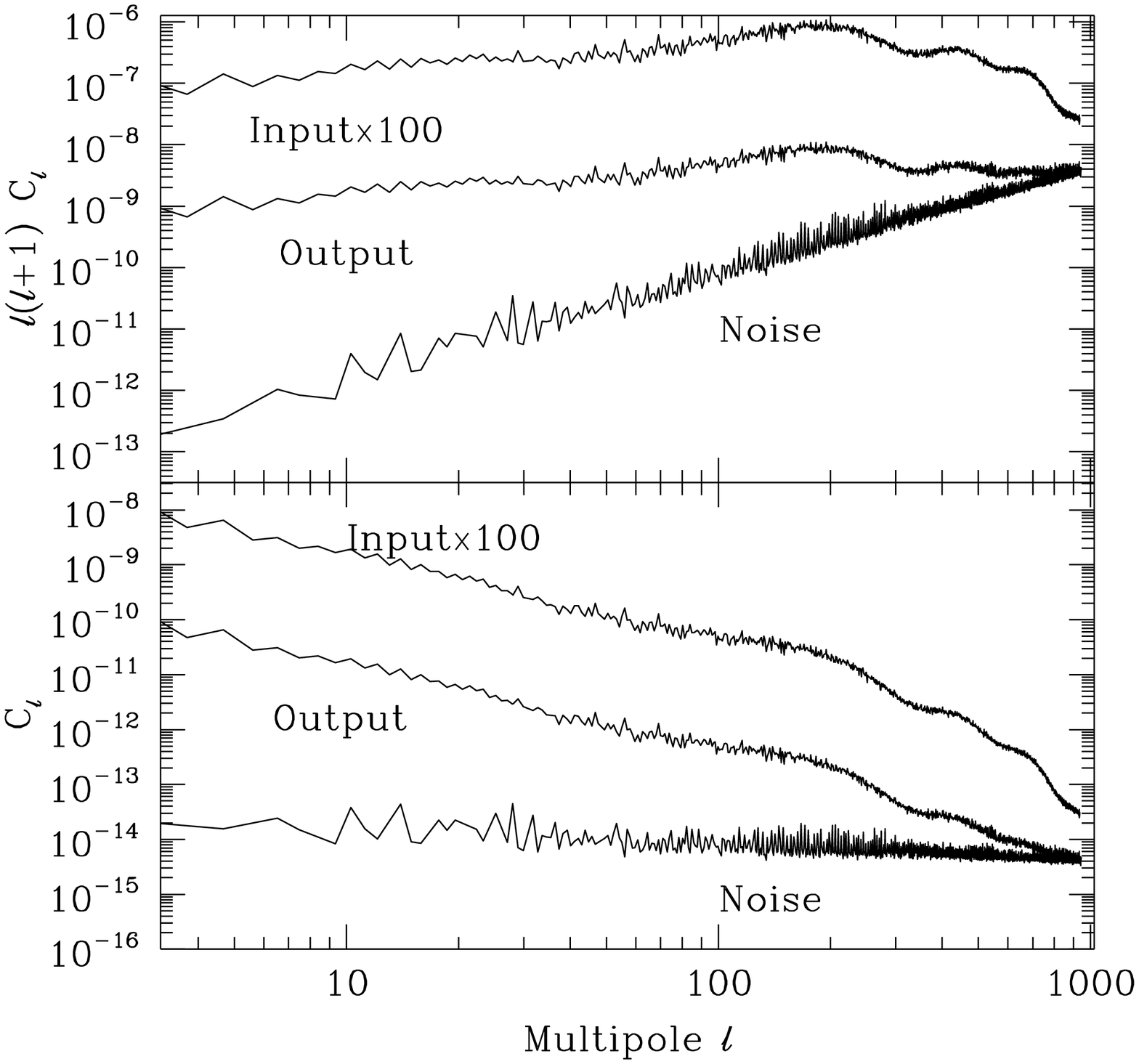}}
\nobreak
\vskip\baselineskip
{\bf Fig.~6a}\ PSI power spectra.
The three curves plotted are from the input, output, and difference images. We
took one realization of a standard CDM sky and observed it with the PSI
spin-chopping scheme.
The top curve is the power spectrum of the (noiseless) input
data, shifted upward to avoid overlap.  Note that the theoretical power spectrum
has been convolved with the experimental beam. The jaggedness of this curve is
just a reflection of the statistical nature of the model.  The middle curve is
the power spectrum fitted to the output image, equivalent to the lower left
panel of Figure~4.  {\it No smoothing has been applied in $\ell$, i.e., every
individual $\ell$ mode is calculated}. The bottom curve is the power spectrum of
the difference image (i.e., output minus input),
equivalent to the lower right panel of
Figure~4. Here the beamsize was taken to be 12\parcm6, with 3\parcm6 pixels, and a
noise level of $24\,\mu$K for a 12\parcm6 resolution element.  Note that the noise
is close to white on all angular scales, and that it only dominates for
$\ell\simgt900$.\quad{\bf Top:}~the familiar way of plotting power spectra, with
$\ell(\ell+1)C_\ell$ on the vertical axis. \quad{\bf Bottom:}~The same plot with
$C_\ell$ on the vertical axis, so that pure white noise appears as a horizontal
line.

\endinsert

\midinsert
\centerline{
\epsfxsize=13cm \epsfysize=13cm \epsfbox{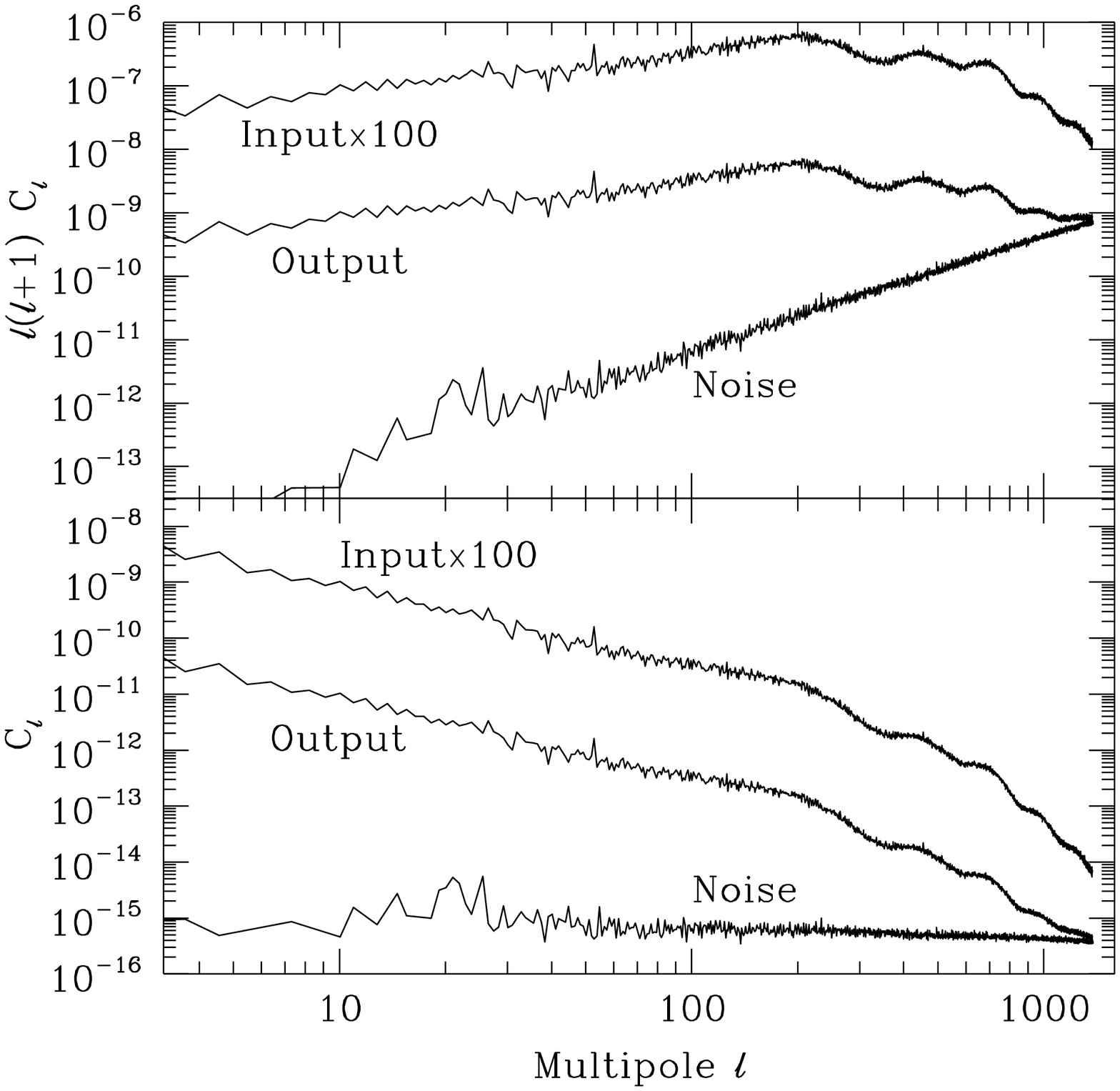}}
\nobreak
\vskip\baselineskip
{\bf Fig.~6b}\ FIRE power spectra.
The same as Figure~6a, but for the parameters of the FIRE experiment.
Here the beamsize was taken to be $8^\prime$, with 2\arcmins25 pixels, and
a noise level of $25\,\mu$K for an $8^\prime$ resolution element.  Note again
that the noise is close to white on all angular scales, and that it only
dominates for $\ell\simgt1400$.\quad{\bf Top:}~the familiar way of plotting
power spectra, with $\ell(\ell+1)C_\ell$ on the vertical axis.
\quad{\bf Bottom:}~The same plot with $C_\ell$ on the vertical axis, so
that pure white noise appears as a horizontal line.

\endinsert

\section CONCLUSIONS\par

PSI and FIRE image the sky in the most direct way possible, scanning it pixel
by pixel.  This strategy, made possible by a combination of technological advances
in detectors, appropriate mission design, and favorable conditions at $L_2$,  has
important technical and scientific advantages over that used by \COBE, in which
differences are measured between widely-spaced parts of the sky.  The technical
advantages include simplicity of the hardware and spacecraft operations, and benign
and extremely stable thermal conditions with the Sun-pointed spin axis.  The
scientific advantages include high angular resolution, high sensitivity,
and the independence of data from different parts of the sky. This last factor
not only eliminates possible aliasing of signal from high foreground regions into
low foreground regions and localizes the effects of transients, but also allows
data to be analyzed as they are taken.  The first slice of the sky can be seen
at the first downlink and analyzed immediately.   Mission  performance and
systematic errors can be investigated almost immediately to microkelvin levels.

Laboratory measurements of the transistor amplifiers and bolometers used by PSI
and FIRE respectively show that their noise characteristics can be modelled
extremely well as a combination of white and $1/f$ noise.  There are no
long-term memory effects, such as those familiar from IRAS, that would result in
persistent offsets after bright objects are crossed.

We have shown both analytically and with extensive simulations of our
experimental procedure that $1/f$ fluctuations on timescales that cannot be 
completely removed by the spacecraft spin are transformed by that spin into a
small increase in near-white noise that introduces {\it no significant
systematic errors}.  Moreover, we have shown that the naive expectation that
$\fknee$ must be much less than the spin frequency is far too restrictive.  PSI and
FIRE can thus realize all of the potential advantages, both scientific and
technical, of the spin-chopping strategy.

\bigskip\bigskip

\noindent {\bf Acknowledgements}
This paper presents the results of one phase of research conducted at the Jet
Propulsion Laboratory, California Institute of Technology, under contract with
the National Aeronautics and Space Administration. We would like to thank all
the members of the teams that were responsible for the development of the PSI
and FIRE mission concepts.  UCSB acknowledges support from NASA grant
NAGW-1062 and NSF grant AST 91-20005 from the Center for Particle Astrophysics.
Further information about the PSI and FIRE missions
can be found on the World Wide Web at these
url's: {\tt http://astrophysics.jpl.nasa.gov/PSI/}
and {\tt http://caseymac.jpl.nasa.gov/fire/fire.html}.
\bigskip
\bigskip


\vskip\parskip
\vskip0.1in
\noindent {\bf References}
\frenchspacing
\parindent=0truept
\baselineskip=12pt

\aref Bennett, C. L., et al., 1994;ApJ;436;423

\aref Boggess, N. W., et al., 1992;ApJ;397;420

\aref Brandt, W. N., Lawrence, C. R., Readhead, A. C. S., Pakianathan, J. N. \&
Fiola, T. M. 1994;ApJ;424;1

\aref Cheng, E. S. \et\ 1995; ApJ; 456; L71

\aref de Bernardis, P. \et\ 1994; ApJ; 422; L33

\apress Gaier, T. \et\ 1996a; IEEE-MTT; submitted

\apre Gaier, T. \et\ 1996b; in preparation

\aref Ganga K., Page, L., Cheng, E., \& Meyer, S. 1994; ApJ; 432; L15

\aref Gundersen, J. O. \et\ 1995; ApJ; 443; L57

\aref Hancock, S. \et\ 1994; Nature; 367; 333

\abook Janssen, M. A. \& Gulkis, S., 1992;{\rm in} The Infrared and
Submillimeter Sky after {\sl COBE};ed. M. Signore \& C. Dupraz, Kluwer,
Dordrecht, p.$\,391$

\aref Lineweaver, C. H. et al., 1994;;ApJ;436;452

\aref Netterfield, C. B. \et\ 1995; ApJ; 445; L69

\apre Piccirillo, L. \et\ 1996; in preparation

\aref Ruhl, J. R. \et\ 1995; ApJ; 453; L1

\apre Seiffert, M. D. \et\ 1996; in preparation

\aref Smoot G. F., et al., 1992;ApJ;396;L1

\apre Tanaka, S. T. \et\ 1996; ApJ; submitted astro-ph/9512067

\aref White, M., Scott, D. \& Silk, J., 1994;ARAA;32;329

\aref Wollack, E. J. \et\ 1993; ApJ; 419; L49

\arep Wright, E. L., Hinshaw, G. \& Bennett, C. L.,
1995;preprint;astro-ph/9510102

\vskip0.2in

\nonfrenchspacing

\end